\documentclass{iau}

\usepackage{array}
\usepackage{multirow}

\usepackage{color}
\usepackage{hyperref}

\title[IAUS291.~~SNRcat] 
{A High-Energy Catalogue\\ of Galactic Supernova Remnants\\ and Pulsar Wind Nebulae} 

\author[Safi-Harb, Ferrand, Matheson]  
{Samar Safi-Harb$^1$, Gilles Ferrand$^2$, Heather Matheson}

\affiliation{Department of Physics \& Astronomy, University of Manitoba,\\ Winnipeg, MB, R3T 2N2, Canada \\ $^1$email: {\tt Samar Safi-Harb <samar@physics.umanitoba.ca>, Canada Research Chair \\
$^2$ CITA National Fellow} \\[\affilskip]}

\pubyear{2012}
\volume{291}  
\jname{\mbox{Neutron Stars and Pulsars: Challenges and Opportunities after 80 years}}
\editors{J. van Leeuwen, ed.} 
\begin{document}

\maketitle

\begin{abstract}
Motivated by the wealth of past, existing, and upcoming X-ray and gamma-ray missions, we have developed the first public database of high-energy observations of all known Galactic Supernova Remnants (SNRs): \url{http://www.physics.umanitoba.ca/snr/SNRcat}. The~ catalogue links to, and complements, other existing related catalogues, including Dave Green's radio SNRs catalogue. We here highlight the features of the high-energy catalogue, including allowing users to filter or sort data for various purposes. The catalogue is currently targeted to Galactic SNR observations with X-ray and gamma-ray missions, and is timely with the upcoming launch of X-ray missions (including Astro-H in 2014). We are currently developing the existing database to include an up-to-date Pulsar Wind Nebulae (PWNe)-dedicated webpage, with the goal to provide a global view of PWNe and their associated neutron stars/pulsars. This extensive database will be useful to both theorists to apply their models or design numerical simulations, and to observers to plan future observations or design new instruments. We welcome input and feedback from the SNR/PWN/neutron stars community.
\end{abstract}


\firstsection 

\section{Objectives}

This work is primarily motivated by the study of particle acceleration: SNRs are believed to be the main production sites of high-energy cosmic rays in the Galaxy, and high-energy observations are particularly important to assess the acceleration of protons. Our~catalogue \cite[(Ferrand \& Safi-Harb 2012)]{Ferrand2012b} was developed with these goals in mind:
\begin{itemize}
\item focus on \textbf{high energies}: Green's catalogue \cite[(2009)]{Green2009b}, mostly based on radio observations, is the reference for the identification and typing of Galactic SNRs. It was used as our base list. We are 
here focused on particle acceleration in SNRs up to TeV or PeV energies, as revealed or hinted for by their broadband X-ray and $\gamma$-ray emission.
\item provide a \textbf{unified view} of SNRs: several observatories already offer dedicated online resources, such as source lists (notably ASCA, Chandra, Fermi, H.E.S.S.) or image galleries (notably ROSAT, Chandra, XMM-Newton). In our catalogue, all observations from the major relevant high-energy observatories are presented together for the first time. Some other websites present all observations in a specific energy domain (notably TeVCat), regardless of the putative emitting objects. Here we offer a complete and broadband (from keV to TeV) view of all Galactic SNRs.
\item be \textbf{up-to-date}: keeping pace with the recent surge in X-ray and $\gamma$-ray observations requires at least weekly updates. 
\item be \textbf{easy to manipulate}: we store the catalogue in a relational database, which allows basic operations such as sorting or filtering.
\end{itemize}

\section{Access} 

Public Access to the catalogue is granted through a dedicated website (located at \url{www.physics.umanitoba.ca/snr/SNRcat}), which provides a pre-defined, simple, almost complete view of the database. 

The main page is the list of all remnants, with each row corresponding to a single object. The first columns of the table describe the SNR (identification, environment, main physical properties), while the last columns summarize the observational status of the remnants for several modern X-ray and gamma-ray instruments. Users are offered various options to sort and filter the table (see online examples).

Clicking on a row opens the full object record in a new page, with more details and references. Again a first table gives all of the properties of the remnant itself, and a second table lists all known high-energy observations of the remnant.

Users are encouraged to send corrections or comments through an online feedback form accessible from any page (\url{www.physics.umanitoba.ca/snr/SNRcat/SNRform.php}).

\section{Statistics} 

Our catalogue provides a summary of our current knowledge of Galactic SNRs from a high-energy perspective. As of 10 October 2012, it contains:
\begin{itemize}
\item \textbf{309 SNR records}: the 274 objects of Green's catalogue as of March 2009, plus 35 objects that were added in light of recent observations. 103 records mention a neutron star (NS) or NS candidate, 85 being identified as a pulsar (PSR). A~pulsar wind nebula (PWN) is detected or suggested in 87 cases (which are not a subset of the former: only 62 SNRs are associated with both a PWN and a NS/PSR). Interaction of the SNR shell with a molecular cloud is reported in 64 cases, with varying levels of confidence. 
\item \textbf{14 records of the sighting of a supernova}, that are referred to by 14 SNRs records (in a non-bijective way, and with varying levels of confidence).
\item \textbf{966 records of high-energy observations} made with 20 observatories, as shown in Table~1. Note that 283 of these are actually non-detections, and that a detected emission might not be coming from the SNR, as seen in Table~2.
\end{itemize}

\section{Perspectives} 

We believe the database will be a very useful tool for the study and modelling of individual remnants as well as to do population studies. It is work in progress, to be updated regularly following new results, in particular from instruments just starting operations (NuSTAR and H.E.S.S. II in 2012) and satellites expected to be soon launched  (eROSITA, ASTROSAT, and Astro-H). In parallel, we plan to extend the catalogue in several directions in the future:
\begin{itemize}
\item \textbf{Objects coverage}: the database includes all SNR types, including PWNe (also referred to as plerions). Our group is currently working on making a dedicated and up-to-date catalogue of PWNe, based on multi-wavelength observations, that will be integrated with the one presented here and to be released in the near future.
\item \textbf{Wavelength coverage}: the database was purposely populated with high-energy observations (with particle acceleration in mind), but the long-term goal is to get a full multi-wavelength view of SNRs, by tighter integration with Green's work in the radio, and by including in-between energy domains (IR, optical, and UV).
\item \textbf{Extragalactic coverage}: following Green's catalogue, the database was deliberately limited to objects located within our Galaxy. It could be extended to extragalactic objects, mostly in the LMC and SMC.
\end{itemize}

\begin{table}[!h]
\newcommand{\llwidth}{1.0cm}
\newcommand{ \lwidth}{1.0cm}
\newcommand{ \rwidth}{3.0cm}
\newcommand{\rrwidth}{3.0cm}

\noindent 
\begin{tabular}{|c|c|c|>{\centering}m{3cm}|>{\centering}m{3cm}|>{\centering}m{3cm}|}
\hline 
\multicolumn{2}{|c|}{\textbf{domain}} & \textbf{instrument} & \textbf{records by instrument} & \multicolumn{2}{c|}{\textbf{records by domain}}\tabularnewline
\hline
\hline 
\multicolumn{2}{|c|}{\multirow{9}{\lwidth}{X-rays}} & ASCA & $111+{\color{white}00}2=113$ & \multicolumn{2}{c|}{\multirow{9}{\rwidth}{$506+{\color{white}0}24=530$}}\tabularnewline
\cline{3-4} 
\multicolumn{2}{|c|}{} & BeppoSAX & ${\color{white}0}17+{\color{white}00}0={\color{white}0}17$ & \multicolumn{2}{c|}{}\tabularnewline
\cline{3-4} 
\multicolumn{2}{|c|}{} & Chandra & $121+{\color{white}00}0=121$ & \multicolumn{2}{c|}{}\tabularnewline
\cline{3-4} 
\multicolumn{2}{|c|}{} & INTEGRAL & ${\color{white}0}21+{\color{white}00}8={\color{white}0}29$ & \multicolumn{2}{c|}{}\tabularnewline
\cline{3-4} 
\multicolumn{2}{|c|}{} & ROSAT & ${\color{white}0}80+{\color{white}00}3={\color{white}0}83$ & \multicolumn{2}{c|}{}\tabularnewline
\cline{3-4} 
\multicolumn{2}{|c|}{} & RXTE & ${\color{white}0}17+{\color{white}00}4={\color{white}0}21$ & \multicolumn{2}{c|}{}\tabularnewline
\cline{3-4} 
\multicolumn{2}{|c|}{} & Suzaku & ${\color{white}0}42+{\color{white}00}1={\color{white}0}43$ & \multicolumn{2}{c|}{}\tabularnewline
\cline{3-4} 
\multicolumn{2}{|c|}{} & SWIFT & ${\color{white}00}8+{\color{white}00}2={\color{white}0}10$ & \multicolumn{2}{c|}{}\tabularnewline
\cline{3-4} 
\multicolumn{2}{|c|}{} & XMM & ${\color{white}0}89+{\color{white}00}4={\color{white}0}93$ & \multicolumn{2}{c|}{}\tabularnewline
\hline 
\multirow{9}{\llwidth}{$\gamma$-rays} & \multirow{2}{\lwidth}{MeV} & COMPTEL & ${\color{white}00}3+{\color{white}00}0={\color{white}00}3$ & \multirow{2}{\rwidth}{${\color{white}00}4+{\color{white}00}2={\color{white}00}6$} & \multirow{9}{\rwidth}{$182+258=440$}\tabularnewline
\cline{3-4} 
 &  & INTEGRAL & ${\color{white}00}1+{\color{white}00}2={\color{white}00}3$ &  & \tabularnewline
\cline{2-5} 
 & \multirow{2}{\lwidth}{GeV} & AGILE & ${\color{white}00}8+{\color{white}00}0={\color{white}00}8$ & \multirow{2}{\rwidth}{${\color{white}0}87+209=296$} & \tabularnewline
\cline{3-4} 
 &  & Fermi & ${\color{white}0}79+209=288$ &  & \tabularnewline
\cline{2-5} 
 & \multirow{8}{\lwidth}{TeV} & ARGO-YBJ & ${\color{white}00}3+{\color{white}00}0={\color{white}00}3$ & \multirow{8}{\rwidth}{${\color{white}0}91+{\color{white}0}47=138$} & \tabularnewline
\cline{3-4} 
 &  & CANGAROO & ${\color{white}00}5+{\color{white}00}7={\color{white}0}12$ &  & \tabularnewline
\cline{3-4} 
 &  & HEGRA & ${\color{white}00}4+{\color{white}00}4={\color{white}00}8$ &  & \tabularnewline
\cline{3-4} 
 &  & H.E.S.S. & ${\color{white}0}50+{\color{white}0}10={\color{white}0}60$ &  & \tabularnewline
\cline{3-4} 
 &  & MAGIC & ${\color{white}00}7+{\color{white}0}12={\color{white}0}19$ &  & \tabularnewline
\cline{3-4} 
 &  & Milagro & ${\color{white}0}10+{\color{white}00}0={\color{white}0}10$ &  & \tabularnewline
\cline{3-4} 
 &  & VERITAS & ${\color{white}0}10+{\color{white}00}6={\color{white}0}16$ &  & \tabularnewline
\cline{3-4} 
 &  & Whipple & ${\color{white}00}2+{\color{white}00}8={\color{white}0}10$ &  & \tabularnewline
\hline
\multicolumn{2}{|c|}{\textbf{ALL}} & \textbf{TOTAL} & $688+282=970$ & \multicolumn{2}{c|}{$688+282=970$}\tabularnewline
\hline
\end{tabular}

\caption{Number of observational records in the database, by energy domain
and by instrument (numbers are the sum of successful observations
and non-detections).}
\end{table}

\begin{table}[!h]
\begin{centering}
\begin{tabular}{|c|c|c|c|c|}
\hline 
 & ejecta / shock & compact object / wind & other (unrelated) & unknown\tabularnewline
\hline
\hline 
X-rays & $207+{\color{white}0}42=249$ & $206+{\color{white}0}60=266$ & ${\color{white}0}17+{\color{white}00}6={\color{white}0}23$ & ${\color{white}0}64$\tabularnewline
\hline 
$\gamma$-rays & ${\color{white}0}33+{\color{white}0}28={\color{white}0}61$ & ${\color{white}0}44+{\color{white}0}22={\color{white}0}66$ & ${\color{white}00}0+{\color{white}00}2={\color{white}00}2$ & ${\color{white}0}60$\tabularnewline
\hline 
TOTAL & $240+{\color{white}0}70=310$ & $250+{\color{white}0}82=332$ & ${\color{white}0}17+{\color{white}00}8={\color{white}0}25$ & $124$\tabularnewline
\hline
\end{tabular}

\caption{Nature of the high-energy emission source for all observational records
in the database (for the first three columns, numbers are the sum
of confident and uncertain identifications). Note that the four columns
are not exclusive.}
\end{centering}

\end{table}

We acknowledge support by the Canada Research Chairs program, the Natural Sciences and Engineering Research Council of Canada, the Canadian Institute for
Theoretical Astrophysics, the Canada Foundation for Innovation, and the Canadian Space Agency.

\end{document}